\documentclass[condensedmatter,article,submit,oneauthor,pdftex,10pt,a4paper]{mdpi} 

\firstpage{1} 
\articlenumber{x}
\doinum{10.3390/------}
\pubvolume{xx}
\pubyear{2016}
\copyrightyear{2016}
\pdfoutput=1

 \theoremstyle{mdpi}
 \newcounter{thm}
 \newcounter{ex}
 \newcounter{re}
 \theoremstyle{mdpidefinition}

\Title{Zero-temperature equation of state of a two-dimensional bosonic quantum fluid with finite-range interaction}

\Author{Andrea Tononi $^{1}$}
\AuthorNames{Andrea Tononi}

\address[1]{%
$^{1}$ \quad Dipartimento di Fisica e Astronomia ``Galileo Galilei'' and CNISM, Universit\`a di Padova, Via Marzolo 8, I-35131 Padova, Italy}

\corres{Correspondence: andrea.tononi@phd.unipd.it}

\abstract{We derive the two-dimensional equation of state 
for a bosonic system of ultracold atoms interacting with a finite-range 
effective interaction. Within a functional integration approach, we employ an 
hydrodynamic parametrization of the bosonic field to calculate 
the superfluid equations of motion and the zero-temperature pressure. 
The ultraviolet divergences, naturally arising from the finite-range 
interaction, are regularized with an improved dimensional regularization 
technique.
}


\begin{document}

%%%%%%%%%%%%%%%%%%%%%%%%%%%%%%%%%%%%%%%%%%
%% Sections that are not mandatory are listed as such. The section titles given are for Articles. Review papers and other article types have a more flexible structure. 

%% Only for the journal Gels: Please place the Experimental Section after the Conclusions

%%%%%%%%%%%%%%%%%%%%%%%%%%%%%%%%%%%%%%%%%%

\section{Introduction}
A fluid perspective on the study of bosonic gases made of ultracold atoms
may find an origin in the pioneering work of Lev Landau \cite{landau1941}, who 
correctly described the superfluid behavior of He-4 observed in 
the experiments \cite{kapitza1938}.
From those years, several technical advances allowed a precise experimental 
control of the atomic gases, culminating in the achievement of Bose-Einstein 
condensation in 1995 \cite{anderson1995,davis1995,hulet1995}.
Despite the rich phenomenology deriving from the variety of different 
trapping potentials \cite{dalfovo1999}, the theoretical and experimental 
study of uniform gases gives a fundamental insight into the 
intrinsic properties of the condensates.
Particularly interesting is the case of two spatial dimensions, in which 
the quantum and thermal fluctuations play a fundamental role 
\cite{merminwagner,hohenberg1967}, justifying the 
necessity of a beyond mean field theory. 
Historical results for a two-dimensional Bose gas were obtained by Schick, 
who calculated the thermodynamics of a gas of hard-spheres \cite{schick1971}, 
improved by Popov derivation of the equation of state for a 
weakly-interacting superfluid \cite{popov1972}.
Recent works provide an extension of Popov approach \cite{pastukhov2018}, while 
the nonuniversal corrections to the equation of state in $D=2$, 
arising from a finite-range interaction between the atoms, have been 
studied \cite{salasnich2017,beane}.
Thank to the tunability of interparticle interactions, 
it is possible investigate the static and dynamical properties of 
homogeneous quantum fluids in $D$-spatial dimensions in regimes 
where finite-range corrections are relevant \cite{braaten,salasnichcappellaro,cappe}.
In this work, we provide an alternative derivation of the zero-temperature 
equation of state by adopting an explicit superfluid parametrization of 
the bosonic field. In particular, we develop an improved dimensional 
regularization technique to regularize the zero-temperature pressure of a 
bosonic quantum fluid, whose particles interact with a finite-range interaction.

%%%%%%%%%%%%%%%%%%%%%%%%%%%%%%%%%%%%%%%%%%
\section{Zero-temperature equation of state of a two-dimensional 
bosonic quantum fluid}
\subsection{Superfluid parametrization of the bosonic field.}
We introduce the Euclidean Lagrangian ${ \cal L }$ of a uniform 
quantum fluid of bosonic particles with mass $m$, described by the 
complex field $\psi(\vec{r},t)$, namely \cite{nagaosa}
\begin{equation}
\label{lagrangian}
{ \cal L } = \bar{\psi}(\vec{r},\tau) \bigg( \hbar\partial_{\tau}-
\frac{\hbar^2 \nabla^2}{2m}-\mu\bigg) \psi(\vec{r},\tau) \\
+ \frac{1}{2}\int d^{D}r \, ' \ |\psi(\vec{r} \, ',\tau)|^{2} \ 
V(|\vec{r}-\vec{r }\, '|) \ |\psi(\vec{r},\tau)|^{2},
\end{equation} 
where $\hbar$ is Planck constant, $\mu$ is the chemical potential 
and we suppose that the 
particles interact with the isotropic two-body potential 
$V(|\vec{r}-\vec{r }\, '|)$. The imaginary time $\tau$ is 
introduced for uniformity with a functional integration approach, 
but real time $t$ can be recovered in any moment performing the 
Wick rotation $\tau \rightarrow it$.
According to the least action principle, the Euler-Lagrange equations 
of the system are obtained as the functional derivative of the 
action $S[\bar{\psi},\psi]$, which reads
\begin{equation}
\label{action}
S[\bar{\psi},\psi] = \int_{0}^{\beta\hbar} d\tau \int_{L^D} d^D r \; 
{ \cal L }
\end{equation}
where $L^D$ is the volume in $D$ dimensions containing the particles 
and $\beta=1/(k_B T)$, with $T$ the absolute temperature and $k_B$ the 
Boltzmann constant. Until the end of this paper, for reasons connected to 
the dimensional regularization of the final results, we will not fix 
explicitly the spatial dimension to $D=2$.
The minimization of the action of Eq. (\ref{action}) gives the Gross-Pitaevski equation 
for the complex field $\psi(\vec{r},\tau)$, which constitutes 
the macroscopic wavefunction of the condensate \cite{leggett}. In this work, however, 
we adopt a superfluid perspective through the following 
phase-amplitude parametrization of the bosonic field
\cite{salasnichlaser}
\begin{equation} \label{phaseamplitude}
\psi(\vec{r},\tau)= \sqrt{\rho(\vec{r},\tau)} \, e^{i \theta(\vec{r},\tau)},
\end{equation}
where $\rho(\vec{r},\tau) = |\psi(\vec{r},\tau)|^{2}$ is a real field 
describing the system local density and $\theta(\vec{r},\tau)$ is the phase field,
which must be included due to the complex nature of the 
order parameter $\psi(\vec{r},\tau)$.
This field transformation allows us to introduce the superfluid velocity 
$\vec{v}_{s}$, which is proportional to the gradient of the phase, namely
\begin{equation}
\vec{v}_{s}= \frac{\hbar}{m} \vec{\nabla}\theta.
\end{equation}
We emphasize that the phase field $\theta(\vec{r},\tau)$ is defined in the compact interval 
$[0, 2\pi]$ and is therefore periodic of $2 \pi$: 
this fact constitutes the origin of many 
topological phenomena in condensed-matter physics. 
Indeed, here we focus on two-dimensional systems, where the singularities of the phase field 
- the vortices - are responsible for the Berezinski-Kosterlitz-Thouless 
(BKT) transition \cite{berezinski1971,kosterlitz1973}.
However, in the following we will study only the zero-temperature properties, for
which the vortex-antivortex phenomenology does not play a fundamental 
role and can be neglected. We will then assume that the domain of 
definition of the phase field $\theta(\vec{r},\tau)$ can be extended 
to $\mathbb{R}$ and that its spatial and time 
derivatives are well defined everywhere.
In this case, the superfluid flow is
irrotational, thus it has zero vorticity
\begin{equation}
\vec{\nabla} \times \vec{v}_{s} =0.
\end{equation}
We now substitute the parametrization of Eq. (\ref{phaseamplitude}) in the 
Lagrangian (\ref{lagrangian}), obtaining
\begin{equation}
\label{lagr2}
{ \cal L } = -\mu \rho  + i\hbar \rho\partial_{\tau}\theta + 
\frac{\hbar^2}{8m \rho}(\nabla \rho)^2 + 
\frac{\hbar^2 \rho}{2m}(\nabla\theta)^2 + \frac{1}{2}\int d^{D}r 
\, ' \ \rho(\vec{r}\,',\tau) \ V(|\vec{r}-\vec{r }\, '|) 
\ \rho(\vec{r},\tau),
\end{equation}
where we omit for simplicity the dependence of the fields 
on their coordinates $(\vec{r},\tau)$.
The minimization of the action (\ref{action}), which now 
becomes a functional of $\rho$ and $\theta$: $S=S[\rho,\theta]$,  
leads to the Euler-Lagrange equations for these fields.
Recovering the real time $t \rightarrow -i \tau$, we get the hydrodynamic 
equations
\begin{equation}
\frac{\partial \rho}{\partial t} + \vec{\nabla} \cdot 
(\rho \vec{v}_{s})=0
\end{equation}
and
\begin{equation} \label{equationmotion}
m \frac{\partial \vec{v}_{s}}{\partial t} + \vec{\nabla} 
\bigg( \frac{1}{2} m v_{s}^2 - \mu + \int d^{D}r \, ' \ 
\rho(\vec{r}\,',\tau) \ 
V(|\vec{r}-\vec{r }\, '|) -\frac{\hbar^{2}}{2m\rho^{1/2}} 
\nabla^{2}(\rho^{1/2}) \bigg)=0,
\end{equation}
which are the continuity equation and the equation of motion of a 
superfluid with velocity $\vec{v_{s}}$ and number density $\rho$. 
Notice that, inside the parenthesis, Eq. (\ref{equationmotion}) 
contains the quantum pressure term:
if the density $\rho$ of the fluid is slowly 
varying and this contribution can be neglected, Eq. 
(\ref{equationmotion}) reduces to the familiar Euler equation 
of motion for an irrotational fluid without viscosity.
For a self-consistent derivation of these equations from the 
Gross-Pitaevski equation we refer the reader to Ref. \cite{coste}.

\subsection{Zero-temperature equation of state.}
We now adopt the superfluid parametrization of Eq. (\ref{phaseamplitude}) 
to derive the zero-temperature equation of state of the quantum fluid, 
namely a relation at $T=0$ between the pressure $P$ and the chemical 
potential $\mu$. For the finite-range interaction, an explicit implementation 
of this relation will be given in the next section.

In the grand canonical ensemble we calculate the pressure of the 
bosonic fluid as $P=-\Omega/L^{D}$, where $\Omega$ is the grand potential 
\begin{equation}
\label{grandpotential}
\Omega=-\frac{1}{\beta} \ln({ \cal Z })
\end{equation}
and ${ \cal Z }$ is the grand canonical partition function, which, 
within a functional integration perspective, can be calculated as
\begin{equation}
\label{partfunction}
{ \cal Z }=\int{{ \cal D} [\rho,\theta] \; e^{-\frac{S[\rho,\theta]}{\hbar}}}.
\end{equation}
To perform the explicit functional integration of the Lagrangian 
(\ref{lagr2}), we rewrite the local density field $\rho(\vec{r},\tau)$ as 
\begin{equation}
\label{parametrizationtheta}
\rho(\vec{r},\tau)=\rho_{0} +\delta\rho(\vec{r},\tau),
\end{equation}
where $\rho_{0}$ is the condensate density of the system in the 
broken-symmetry phase and $\delta \rho(\vec{r},\tau)$ is a real field 
describing the density fluctuations.

We substitute the field transformation (\ref{parametrizationtheta}) 
in the Lagrangian of Eq. (\ref{lagr2}), obtaining
\begin{align}
\begin{split}
\label{lagrrhotheta}
{ \cal L } =& -\mu \rho_{0} - \mu \, \delta\rho + i\hbar \rho_{0}
\partial_{\tau}\theta + \frac{\hbar^2}{8m \rho_{0}}(\nabla\delta \rho)^2 
+ \frac{\hbar^2 \rho_{0}}{2m}(\nabla\theta)^2 + \\ 
&\frac{1}{2}\int d^{D}r' \  V(|\vec{r}-\vec{r }\, '|) 
(\rho_{0}^2 + \delta \rho(\vec{r},\tau) + \delta \rho(\vec{r} \, ',\tau) 
+ \rho(\vec{r},\tau) \delta \rho(\vec{r} \, ',\tau)),
\end{split}
\end{align}
where we keep only terms up to second order in the fluctuation 
fields $\delta \rho(\vec{r},\tau)$ and $\theta(\vec{r},\tau)$, 
thus making a Gaussian (one-loop) approximation.

Considering the Lagrangian of Eq. (\ref{lagrrhotheta}) inside the action $S$,
which now becomes the functional $S=S[\delta\rho,\theta]$,
it is particularly convenient to express it
in terms of the Fourier series of the fluctuation fields, namely
\begin{align}
\delta \rho(\vec{r},\tau)&=\frac{1}{\sqrt[]{L^D}}\sum_{\vec{k} \, 
\omega_{n}} e^{i\vec{k}\cdot\vec{r}}e^{-i\omega_{n}\tau} \, \delta 
\rho(\vec{k},\omega_{n}) \nonumber \\ \theta(\vec{r},\tau)
&=\frac{1}{\sqrt[]{L^D}}\sum_{\vec{k} \, \omega_{n}} 
e^{i\vec{k}\cdot\vec{r}}e^{-i\omega_{n}\tau} \, \theta(\vec{k},\omega_{n}) 
\nonumber \\ \delta \rho(\vec{k},\omega_{n})
&=\frac{1}{\beta\hbar \ \sqrt[]{L^D}}\int_{0}^{\beta\hbar} 
d\tau \int_{L^D} d^D r \, e^{-i\vec{k}\cdot\vec{r}}e^{i\omega_{n}\tau}
\, \delta \rho(\vec{r},\tau) \nonumber \\ 
\theta(\vec{k},\omega_{n})&=\frac{1}{\beta\hbar \ 
\sqrt[]{L^D}}\int_{0}^{\beta\hbar} d\tau 
\int_{L^D} d^D r \, e^{-i\vec{k}\cdot\vec{r}}
e^{i\omega_{n}\tau} \, \theta(\vec{r},\tau), 
\end{align}
where $\omega_n=2\pi n/(\beta \hbar)$ are the bosonic Matsubara frequencies.
Notice that, since we are supposing that the phase field 
$\theta(\vec{r},\tau)$ is defined on $\mathbb{R}$, 
its Fourier components are non-numerable and can assume continuous values, 
thus they can be treated like ordinary functional integral variables.
The action in the Fourier space is obtained by simply substituting 
these expressions in $S$ and using the definition of 
the $D+1$-dimensional delta function.
Moreover, we also substitute the Fourier series $\tilde{V}(k)$ of 
the real space interaction potential and we define with 
$g_0$ the zero-range interaction strength $g_0=\tilde{V}(k=0)$.
In this way, the action can be rewritten as the sum of two contributions
\begin{equation}
S = S_{0} + S_{g}.
\end{equation}
The first is the action of the homogeneous system $S_{0}$, namely
\begin{equation}
S_{0}=\beta\hbar L^D \big(  -\mu \rho_{0} + \frac{1}{2}g_{0} \rho_{0}^2 \big),
\end{equation}
which does not depend on the functional integration variables:
using Eqs. (\ref{grandpotential}) and (\ref{partfunction}) one 
can employ $S_0$ to calculate $\Omega_{0}$, the 
mean field contribution to the grand potential 
\begin{equation} \label{omega0}
\Omega_{0} = \big( -\mu \rho_{0} + \frac{1}{2}g_{0} \rho_{0}^2  \big) L^D.
\end{equation}
The second contribution to the action $S$ is the Gaussian action $S_{g}$, 
which is given by
\begin{equation}
S_{g} = \beta\hbar\sum_{\vec{k} \, \omega_{n}} 
\bigg[\frac{\hbar^2k^2\rho_{0}}{2m}\theta(k)\theta(-k) 
+ \bigg(\frac{\hbar^2k^2}{8m\rho_{0}} + 
\frac{\tilde{V}(k)}{2}\bigg)\delta \rho(k)\delta \rho(-k) 
+ \hbar\omega_{n}\theta(k)\delta\rho(-k)\bigg],
\end{equation}
where, for simplicity of notation, we define 
$\delta \rho(\pm k)=\delta \rho(\pm \vec{k},\pm \omega_n)$ 
and $\theta(\pm k)=\theta(\pm \vec{k},\pm \omega_n)$.
Since $S_{g}$ is quadratic in the fluctuation fields $\delta \rho(k)$ 
and $\theta(k)$, one can rewrite it in the following matricial form
\begin{equation}
S_{g} = \frac{\hbar}{2}\sum_{\vec{k} \, \omega_{n}}
\begin{pmatrix}
\theta(k) & \theta(-k) & \delta \rho(k) & \delta \rho(-k)
\end{pmatrix}\textbf{M}(k)
\begin{pmatrix}
\theta(k) \\
\theta(-k) \\
\delta \rho(k) \\
\delta \rho(-k)
\end{pmatrix},
\end{equation}
where $\textbf{M}(k)$, the inverse of the propagator, is the $4 \times 4$ matrix
\begin{equation}
\textbf{M}(k) = \beta 
\left( \begin{array}{c c c c} 
0 & \frac{\hbar^2k^2\rho_{0}}{m} & 0 & \hbar\omega_{n} \\ 
\frac{\hbar^2k^2\rho_{0}}{m} & 0 & -\hbar\omega_{n} & 0 \\ 
0 & -\hbar\omega_{n} & 0 & \frac{\hbar^2k^2}{4 m\rho_{0}} 
+ \tilde{V}(k) \\ 
\hbar\omega_{n} & 0 & \frac{\hbar^2k^2}{4 m\rho_{0}} 
+ \tilde{V}(k) & 0 \\ 
\end{array} \right).
\end{equation}
The functional integral of the real fluctuation fields 
$\theta(k)$ and $\delta \rho(k)$ can be performed explicitly \cite{altland2010},
obtaining the corresponding Gaussian grand canonical partition 
function $\mathcal{Z}_{g}$ as
\begin{equation}
{ \cal Z }_{g} = \prod_{\substack{\vec{k} \, \omega_{n} \\ k_{z}>0}} 
[\det \textbf{M}(k)]^{-1/2},
\end{equation}
which, considering the definition of the grand potential of Eq. 
(\ref{grandpotential}), leads to the Gaussian contribution to the 
grand potential 
\begin{equation}
\label{omegag}
\Omega_{g} = \frac{1}{2 \beta} \sum_{\vec{k} \, \omega_{n}} 
\ln[\beta^2(\hbar^2\omega_{n}^2+E_{k}^2)].
\end{equation}
Here, we find the gapless excitation spectrum $E_{k}$ 
of the quantum fluid in the form
\begin{equation}
E_{k} = \sqrt[]{\frac{\hbar^2 k^2}{2m} \bigg( \frac{\hbar^2 k^2}{2m} 
+ 2 \rho_{0} \tilde{V}(k)  \bigg) },
\end{equation}
where, within a perturbative approach, $\rho_0$ is determined by the 
saddle point condition $\partial\Omega_{0}/\partial \rho_{0} = 0$, 
which leads to 
\begin{equation}\label{spc}
\rho_{0} = \frac{\mu}{g_{0}}
\end{equation}
and whose substitution in the excitation spectrum gives $E_{k}^B$, the 
renowned Bogoliubov spectrum \cite{bogoliubov1947}
\begin{equation}
E_{k}^B = \sqrt[]{\frac{\hbar^2 k^2}{2m} \bigg( \frac{\hbar^2 k^2}{2m} 
+ 2\mu \frac{\tilde{V}(k)}{g_0} \bigg) }.
\end{equation}\label{bogoliubovspectrum}
The sum over the Matsubara frequencies $\omega_n$ in the Gaussian 
grand potential of Eq. (\ref{omegag}) can be performed 
according to the prescriptions described in the Appendix, 
obtaining the grand potential as the sum of three contributions
\begin{equation} \label{gptotal}
\Omega = \Omega_{0} + \Omega_{g}^{(0)} + \Omega_{g}^{(T)},
\end{equation}
where $\Omega_{0}=-L^D \mu^2/(2g_0)$ due to Eqs. (\ref{omega0}) and (\ref{spc}), 
and 
\begin{equation}
\label{omegag0}
\Omega_{g}^{(0)}= \frac{1}{2} \sum_{\vec{k}} E_{k}^B
\end{equation}
is the zero-temperature Gaussian grand potential encoding quantum fluctuations, while
\begin{equation}
\label{omegagT}
\Omega_{g}^{(T)}= \frac{1}{\beta} \sum_{\vec{k}} \ln(1-e^{-\beta E_{\vec{k}}^B})
\end{equation}
is the finite-temperature Gaussian grand potential, encoding thermal fluctuations.
Finally, we explicitly write the zero-temperature equation of state, namely we 
calculate the pressure as the opposite of the grand potential of Eq. (\ref{gptotal}) 
at $T=0$:
\begin{equation}
P(\mu,T=0)=\frac{\mu^2}{2 g_0}- \frac{1}{2 L^D} \sum_{\vec{k}} E_{k}^B.
\label{eqstate0}
\end{equation}
In the thermodynamic limit of $L \rightarrow \infty$, the sum over 
$\vec{k}$ can be rewritten as a $D$-dimensional integral in 
momentum space $(2 \pi/L)^{D} \sum_{\vec{k}} = \int d^{D} k$, and, substituting again the Bogoliubov spectrum 
(\ref{bogoliubovspectrum}), the equation of state becomes 
\begin{equation}
P(\mu,T=0)=\frac{\mu^2}{2 g_0}- \frac{1}{2} \int \frac{d^{D} k}{(2 \pi)^{D}} 
\; \sqrt[]{\frac{\hbar^2 k^2}{2m} \bigg( \frac{\hbar^2 k^2}{2m} 
+ 2\mu \frac{\tilde{V}(k)}{g_0} \bigg) },
\label{eqstate}
\end{equation}
where the integral can be calculated after the explicit choice of $\tilde{V}(k)$.

\subsection{Explicit implementation for finite-range interaction.}
We now provide an explicit implementation 
of the zero-temperature equation of state (\ref{eqstate}) for a bosonic quantum 
fluid of particles interacting with the finite-range effective 
interaction
\begin{equation} \label{finiterange}
\tilde{V}(k)=g_0 +g_2 k^2,
\end{equation}
where $g_0=\tilde{V}(k=0)$ is the usual zero-range interaction coupling, and 
\begin{equation}
g_2= \frac{1}{2} \int d^2 r \, r^2 \, V(|\vec{r}|)
\end{equation}
is the first nonzero correction in the gradient expansion of an isotropic 
interaction potential $V(|\vec{r}|)$. At zero temperature, 
we expect that the finite-range 
corrections to the equation of state are detectable, 
but small with respect to the zero-range result of Ref. \cite{salasnich2016}.
By using scattering theory in 
two spatial dimensions, these couplings can be linked with 
the s-wave scattering length $a_s$ and the characteristic 
range $R$ of the real interatomic two-body interaction \cite{tononi,astrakharchik2009,salasnich2017}
\begin{equation}
g_0= \frac{4 \pi \hbar^2}{m |\ln(n a_s^2)|}, \qquad g_2= \frac{\pi \hbar^2 R^2}{m |\ln(n a_s^2)|},
\end{equation}
where $n$ is the number density of the system in $D=2$.

The equation of state (\ref{eqstate}) becomes, with the finite-range interaction of 
Eq. (\ref{finiterange})
\begin{equation}
P(\mu,T=0)=\frac{\mu^2}{2 g_0} + P_{g}^{(0)},
\end{equation}
where we define the zero temperature Gaussian pressure $P_{g}^{(0)}$ as
\begin{equation}
P_{g}^{(0)}=- \frac{1}{2} \int \frac{d^{D} k}{(2 \pi)^{D}}
 \; \sqrt[]{\frac{\hbar^2k^2}{2m}
\bigg(\frac{\hbar^2k^2}{2m} \lambda +2 \mu \bigg)},
\end{equation}
with
\begin{equation}
\label{lambda}
\lambda = 1+ \frac{4m}{\hbar^2} \frac{\mu}{g_{0}} g_{2}.
\end{equation}
Since the integrand function depends only on the modulus of 
the momentum $|\vec{k}|$, we rewrite the integral in $P_{g}^{(0)}$ 
using $D$-dimensional spherical coordinates, namely
\begin{equation} 
P_{g}^{(0)}  = -\frac{S_{D}}{2 (2 \pi)^{D}} 
\int_{0}^{+\infty} dk \, k^{D-1} \ \sqrt[]{\frac{\hbar^2k^2}{2m}
\bigg(\frac{\hbar^2k^2}{2m} \lambda +2 \mu \bigg)},
\end{equation}
where $S_{D}=2 \pi^{D/2}/\Gamma[D/2]$ is the solid angle in $D$-dimensions and 
$\Gamma[D/2]$ is the Euler Gamma function.
In order to integrate this equation, we introduce the adimensional
variable $t= \hbar^2k^2 \lambda/(4 m \mu)$, obtaining 
\begin{equation} \label{pg0divergent}
P_{g}^{(0)} = -\frac{\mu}{{\lambda}^{1/2} \Gamma[D/2]}
\bigg( \frac{m \mu}{\pi \hbar^{2} 
\lambda} \bigg)^{D/2}  \  
\int_{0}^{+\infty} dt \ t^{\frac{D-1}{2}} (1+t)^{1/2}.
\end{equation}
As a consequence of the substitution of the real interatomic 
potential with an effective interaction, 
the zero-temperature Gaussian pressure $P_{g}^{(0)}$ is ultraviolet divergent.
In our framework, an efficient way to regularize $P_{g}^{(0)}$ 
is constituted by the technique of 
dimensional regularization \cite{thooft1972}. 
The basic idea of this approach is to rewrite a diverging integral 
in terms of the Euler beta and gamma  
functions, whose integral representation for $x,y,z>0$ is given by 
\begin{align}
\label{betafunction}
B(x,y)=\int_{0}^{+\infty} dt \ \frac{t^{x-1}}{(1+t)^{x+y}},
\end{align}
\begin{align}
\Gamma(z) = \int_{0}^{+\infty} dt \ t^{z-1} \ e^{-z}.
\end{align}
Thank to the properties $B(x,y)=\Gamma(x)\Gamma(y)/\Gamma(x+y)$ 
and $\Gamma[z+1]=z \, \Gamma[z]$, 
one can extend the domain of definition of the gamma and beta functions 
by analytic continuation of their arguments $x,y,z$ also to negative values, 
which usually appear in many physical problems.
However, despite this dimensional regularization procedure 
can be successfully used to regularize many 
ultraviolet diverging integrals, in our peculiar two-dimensional case 
the procedure described above would lead to a result containing the 
gamma function evaluated for negative {\it integer} values, which is 
again a diverging quantity.
To avoid this residual divergence, we extend the dimension of the system to the 
complex value $\mathcal{D} = D - \varepsilon$, and we formally perform the 
integration of Eq. (\ref{pg0divergent}). We obtain 
\begin{equation}
\label{finiterangepressureregularized}
P_{g}^{(0)} = \frac{\kappa^\varepsilon}{2} \bigg( \frac{\mu}{\pi \lambda} \bigg)^{(\mathcal{D}+1)/2}
 \bigg( \frac{m}{\hbar^2} \bigg)^{\mathcal{D}/2}
    \frac{\Gamma[(D - \varepsilon+1)/2] \ 
    \Gamma[(\varepsilon -D -2)/2]}{\Gamma[(D - \varepsilon)/2]},
\end{equation}
in which the wavevector $\kappa$ is introduced for dimensional reasons.
Notice how in $D=2$ and for $\varepsilon=0$ the Gaussian pressure is 
still divergent. To regularize it, we rely on the following small-$\varepsilon$ 
expansion of the gamma function \cite{kleinert2001}
\begin{equation}
\label{Gammanepsilon}
\Gamma (-n + \varepsilon) = \frac{(-1)^{n}}{n!} 
\bigg[ \frac{1}{\varepsilon} + \Psi(n+1) + 
\frac{\varepsilon}{2} \bigg( \frac{\pi^2}{3} + 
{\Psi(n+1)}^{2} - \Psi'(n+1) \bigg) + o(\varepsilon^2) 
 \bigg],
\end{equation}
where $\Psi(n+1)$ is the digamma function and $\Psi'(n+1)$ is its 
derivative. Moreover, we express the exponentiation of a generic 
coefficient $x^{\varepsilon}$ for $\varepsilon \to 0$ as
\begin{equation}
\label{epsilonexpansion}
x^{\varepsilon}  = \exp( \varepsilon \ln(x)) 
\sim_{\varepsilon \rightarrow 0} \ 1 + \varepsilon 
\ln(x) + o(\varepsilon^2).
\end{equation}
With this recipe, the Gaussian pressure $P_{g}^{(0)}$ in $D=2$ gives 
\begin{equation}
P_{g}^{(0)} = \frac{m \mu^{2}}{2 \pi^{3/2} 
{\hbar^{2} \lambda}^{3/2}} \ 
\bigg[\frac{\pi^{1/2}}{2} \frac{1}{\varepsilon} + 
\frac{\pi^{1/2}}{8} (\ln (16) - 2 \gamma -1) + 
\frac{\pi^{1/2}}{4} \ln \bigg( \frac{\pi \lambda \hbar^{2}
 \kappa^{2}}{m \mu} \bigg) + o(\varepsilon) \bigg],
\end{equation}
where $\gamma \approx 0.55722$ is the Euler-Mascheroni 
constant.
Finally, we delete the $o(\varepsilon^{-1})$ divergence in the square bracket 
\cite{zeidler2009} and we rewrite the zero-temperature equation 
of state $P(\mu,T=0)$ of Eq. (\ref{eqstate}) as
\begin{equation}
\label{pressure0Tfiniterange}
P(\mu,T=0) = \frac{m \mu^{2}}{8 \pi \hbar^{2} 
{\lambda}^{3/2}} \ \bigg[ \ln \bigg( \frac{\epsilon_{0}}
{\mu}\lambda \bigg) - \frac{1}{2} \bigg],
\end{equation}
where we define the energy cutoff $\epsilon_{0}$ as
\begin{equation}
\epsilon_{0} = \frac{4 \pi \hbar^{2} \kappa^{2}  }
{m \exp(\gamma - \frac{4 \pi \hbar^{2} {\lambda}^{3/2}}{m g_{0}}) }.
\end{equation}
The equation of state (\ref{pressure0Tfiniterange}) improves the one derived 
for bosons with a zero-range interaction \cite{popov1972} by Popov, 
whose result can be reproduced by setting $\lambda=1$, i.e. $g_2=0$.
We emphasize that, with a precise tuning of the interparticle interaction 
(see Ref. \cite{salasnich2017} for a detailed discussion), the finite-range corrections derived 
within our Gaussian approximation become larger than the zero-range 
beyond-Gaussian ones obtained by Mora and Castin \cite{moracastin}.
For weakly-interacting bosons with $n a_s^2 \ll 1$, where $a_s$ is
the two-dimensional s-wave scattering length, we expect that the nonuniversal 
corrections of Eq. (\ref{pressure0Tfiniterange}) arise for $R \ge a_s$, 
where $R$ is the characteristic range of the interaction. 
In this intermediate regime the neglection of higher order terms in the gradient 
expansion of Eq. (\ref{finiterange}) is justified but, at the same time, the 
finite-range contributions are of comparable size to the zero-range ones.

\section{Conclusions}
In this work we derive the two-dimensional zero-temperature equation of state 
for a bosonic quantum fluid with a generic isotropic interaction.
The superfluid perspective is emphasized by performing the Gaussian 
functional integration within a phase-amplitude parametrization of 
the complex order parameter.
For a system with zero-range interaction, we reproduce the classical 
result by Popov. Nonetheless, we apply a novel dimensional 
regularization recipe to reproduce the nonuniversal corrections 
for a finite-range interaction potential.
We expect that, with a fine tuning of the experimental interaction 
parameters, the finite-range correction produce sizable corrections 
to the thermodynamics of the weakly-interacting superfluid.
Our derivation of the zero-temperature equation of state is valid 
also for other interparticle interactions. 
In particular, the previous results can be extended for a quasi-two-dimensional 
system of dipolar bosons whose polarization direction is 
perpendicular to the plane of confinement.
For a generic orientation, however, it is necessary to consider the 
dependence of the interaction on the in-plane angle between the 
particles and to include it consistently in the dimensional 
regularization procedure.

%%%%%%%%%%%%%%%%%%%%%%%%%%%%%%%%%%%%%%%%%%
\section{Appendix}
\label{matsubarasumappendix}
We illustrate here the procedure to calculate the summation 
over the bosonic Matsubara frequencies $\omega_{n}$, which 
are defined as
\begin{equation}
\omega_{n}=\frac{2 \pi n}{\beta \hbar},
\end{equation} 
where $n \in \mathbb{Z}$ are integer numbers. The most common 
sum that one has to perform is in the form
\begin{equation}
\label{matsubarasum}
I[\xi_{\vec{k}}]= \frac{1}{2 \beta}  \sum_{n=-\infty}^{+\infty} 
\ln [\beta^{2}(\hbar^{2}\omega_{n}^{2}+ \xi_{\vec{k}}^{2})].
\end{equation}
Using the properties of the logarithm and considering that the 
summation involves all $n \in \mathbb{Z}$ integers, both 
positive and negative, $I[\xi_{\vec{k}}]$ can also be 
rewritten in the useful form
\begin{equation}
I[\xi_{\vec{k}}] = \frac{1}{\beta} \sum_{n=-\infty}^{+\infty} 
\ln [\beta (-i \hbar \omega_{n}+ \xi_{\vec{k}})].
\end{equation}
Taking the derivative of $I[\xi_{\vec{k}}]$ with respect to 
$\xi_{\vec{k}}$ in the Eq. (\ref{matsubarasum}) we get
\begin{equation}
\label{matsubarasum1}
\frac{\partial I[\xi_{\vec{k}}]}{\partial \xi_{\vec{k}}} = 
\frac{1}{\beta}  \sum_{n=-\infty}^{+\infty}  
\frac{\xi_{\vec{k}}}{\hbar^{2}\omega_{n}^{2}+ \xi_{\vec{k}}^{2}}. 
\end{equation}
In the zero temperature limit, the difference
\begin{equation}
\Delta\omega=\omega_{n}-\omega_{n-1}=\frac{2 \pi}{\beta \hbar} 
\xrightarrow[\beta >> 1]{} \ d \omega
\end{equation}
becomes infinitesimal and we can substitute the sum over 
$n$ with an integral over $\omega$, obtaining
\begin{equation}
\frac{\partial I[\xi_{\vec{k}}]}{\partial \xi_{\vec{k}}} = 
\frac{1}{\beta}  \int_{-\infty}^{+\infty} d \omega \ 
\frac{\beta\hbar}{2 \pi} \ \frac{\xi_{\vec{k}}}{\hbar^{2}\omega^{2}
+ \xi_{\vec{k}}^{2}} = \frac{1}{2},
\end{equation}
which is the zero-temperature contribution to $I[\xi_{\vec{k}}]$. 
If the temperature is relatively low, but non-zero, 
we cannot substitute the sum in Eq. (\ref{matsubarasum1}) 
with an integral, but we can rewrite it as
\begin{equation}
\frac{\partial I[\xi_{\vec{k}}]}{\partial \xi_{\vec{k}}} = 
\frac{\beta \xi_{\vec{k}}}{(2 \pi)^{2}}  \sum_{n=-\infty}^{+\infty}  
\frac{1}{n^{2}+ \big(\frac{\beta \xi_{\vec{k}}}{2 \pi}\big)^{2}} 
\end{equation}
and, using the identity 
\begin{equation}
\sum_{n=0}^{+\infty}  \frac{1}{n^{2}+ a^{2}} = 
\frac{1+\pi a \ \coth(\pi a)}{2 a^{2}},
\end{equation}
we obtain
\begin{equation}
\frac{\partial I[\xi_{\vec{k}}]}{\partial \xi_{\vec{k}}} = 
\frac{1}{2} \coth \bigg(\frac{\beta\xi_{\vec{k}}}{2} \bigg) = 
\frac{1}{2} + \frac{1}{e^{\beta\xi_{\vec{k}}}-1}.
\end{equation}
We integrate this equation on $\xi_{\vec{k}}$ and, setting 
the arbitrary constant resulting from the indefinite integral 
to zero (it is not dependent on physical parameters), we 
finally obtain the result of the summation over the Matsubara frequencies
\begin{equation}
\label{matsubarasumformula}
I[\xi_{\vec{k}}] = \frac{\xi_{\vec{k}}}{2} + 
\frac{1}{\beta} \ln(1-e^{-\beta\xi_{\vec{k}}}),
\end{equation}
which is used in this article to obtain Eq. (\ref{omegag}).

%%%%%%%%%%%%%%%%%%%%%%%%%%%%%%%%%%%%%%%%%%
\acknowledgments{The author thanks Luca Salasnich and Alberto Cappellaro for useful discussions and suggestions.}

%%%%%%%%%%%%%%%%%%%%%%%%%%%%%%%%%%%%%%%%%%
\conflictofinterests{``The author declares no conflict of interest.''}

%%%%%%%%%%%%%%%%%%%%%%%%%%%%%%%%%%%%%%%%%%
\bibliographystyle{mdpi}

%=====================================
% References, variant A: internal bibliography
%=====================================
\renewcommand\bibname{References}

%%%%%%%%%%%%%%%%%%%%%%%%%%%%%%%%%%%%%%%%%%
\end{document}